\documentclass[sigconf]{acmart}

\AtBeginDocument{%
  }

\copyrightyear{2025}
\acmYear{2025}
\setcopyright{rightsretained}
\acmConference[CHI EA '25]{Extended Abstracts of the CHI Conference on Human Factors in Computing Systems}{April 26-May 1, 2025}{Yokohama, Japan}
\acmBooktitle{Extended Abstracts of the CHI Conference on Human Factors in Computing Systems (CHI EA '25), April 26-May 1, 2025, Yokohama, Japan}\acmDOI{10.1145/3706599.3719714}
\acmISBN{979-8-4007-1395-8/2025/04}

\begin{document}


\title{Towards Understanding the Use of MLLM-Enabled Applications for Visual Interpretation by Blind and Low Vision People}

\author{Ricardo E. Gonzalez Penuela}
\affiliation{%
  \institution{Cornell University}
  \city{New York}
  \country{USA}}
\email{reg258@cornell.edu}


\author{Ruiying Hu}
\email{rh692@cornell.edu}
\affiliation{%
  \institution{Cornell Tech}
  \city{New York}
  \state{New York}
  \country{USA}
}

\author{Sharon Lin}
\email{syl55@cornell.edu}
\affiliation{%
  \institution{Cornell Tech}
  \city{New York}
  \state{New York}
  \country{USA}
}

\author{Tanisha Shende}
\email{tshende@oberlin.edu}
\affiliation{%
  \institution{Oberlin College}
  \city{Oberlin}
  \state{Ohio}
  \country{USA}
}

\author{Shiri Azenkot}
\email{shiri.azenkot@cornell.edu}
\affiliation{%
  \institution{Cornell Tech}
  \city{New York}
  \state{New York}
  \country{USA}
}

\renewcommand{\shortauthors}{Gonzalez et al.}

\begin{abstract}
Blind and Low Vision (BLV) people have adopted AI-powered visual interpretation applications to address their daily needs. While these applications have been helpful, prior work has found that users remain unsatisfied by their frequent errors. Recently, multimodal large language models (MLLMs) have been integrated into visual interpretation applications, and they show promise for more descriptive visual interpretations. However, it is still unknown how this advancement has changed people’s use of these applications. To address this gap, we conducted a two-week diary study in which 20 BLV people used an MLLM-enabled visual interpretation application we developed, and we collected 553 entries. In this paper, we report a preliminary analysis of 60 diary entries from 6 participants. We found that participants considered the application’s visual interpretations trustworthy (mean 3.75 out of 5) and satisfying (mean 4.15 out of 5). Moreover, participants trusted our application in high-stakes scenarios, such as receiving medical dosage advice. We discuss our plan to complete our analysis to inform the design of future MLLM-enabled visual interpretation systems.

\end{abstract}

\ccsdesc[500]{Human-centered computing~Empirical studies in accessibility}
\ccsdesc[500]{Human-centered computing~Empirical studies in HCI}
\ccsdesc[500]{Human-centered computing~Field studies}
\ccsdesc[300]{Human-centered computing~Accessibility technologies}

\keywords{Accessibility, BLV, Blind People, Low Vision People, AI, Artificial Intelligence, LLM, Large language models, Diary Study, Conversations, MLLM, Multimodal models, VQA, Visual Question Answering}

\maketitle

\section{Introduction}

Perceiving visual information is challenging for Blind and Low Vision (BLV) people, leading to difficulties with daily activities like traveling, shopping, and navigating unfamiliar spaces \cite{kamikuboShopping, jeamwatthanachai2019indoor}. 

AI-powered visual interpretation applications, such as SeeingAI \cite{seeingAI}, have become essential tools for supporting Blind and Low Vision (BLV) people in accessing visual information. Users can take a photo to identify objects or describe a scene. For example, a Blind parent can take a photo while cooking for their children to identify a can of soup or a Low Vision traveler can take a photo to identify their hotel room number. 

Although these applications are helpful for many use cases, researchers have found that BLV users are not very satisfied with the AI-generated visual interpretations they produce \cite{gonzalez2024usecases, kupferstein2020understanding}. For instance, we conducted a diary study to investigate BLV people's use of visual interpretation applications before the implementation of multimodal large language models (MLLMs) into mainstream applications \cite{gonzalez2024usecases}. In our study, we found that users’ satisfaction was low when trying to build an understanding of their visual surroundings and when trying to identify visual features of qualities like the color of a shirt (mean=2.58, and 2.23, respectively, out of a 5-point scale). 

More recently, SeeingAI, Be My AI, and other applications have incorporated MLLMs to improve the quality of their visual interpretation systems \cite{be_my_ai_blog, applevis2025seeingai}. Unlike prior computer vision models, these MLLMs can make sense of multiple user inputs including photos and user questions to provide more detailed visual descriptions \cite{openai_gpt4v_system_card}. Naturally, BLV people who use SeeingAI and Be My Eyes now enjoy access to MLLM-enabled visual interpretations. Recent research also found that BLV people are adopting MLLM-enabled tools because of their usability and usefulness to support with their work \cite{UseAi_das, aigeneratedimg_bennett}. For example, Adnin et al. interviewed BLV people to understand the accessibility of MLLM-enabled tools and found that, while useful for creative content generation, Blind users face significant accessibility barriers navigating the interfaces of ChatGPT, Gemini, and other MLLM-enabled tools \cite{UseAi_das}. Similarly, recent work shows that MLLMs can produce well-written, long-form answers to user-generated visual questions; but their answers often contain factual errors (that go undetected by BLV people) \cite{chen2025fully,longvqa_huh}. Moreover, the questions they evaluated do not reflect current BLV people’s experiences with MLLM-enabled visual interpretation systems \cite{chen2025fully,longvqa_huh}.

Thus, while prior work has explored BLV people’s initial engagement with MLLM-enabled applications and has validated the potential of MLLMs to produce answer to visual questions, we do not have a clear understanding of how BLV users interact with MLLM-enabled visual interpretation applications, and whether they meet BLV people's needs.  In other words, it remains unclear how recent advancement of AI has affected BLV people's experience with such applications.

To address this gap, our work seeks to answer the following research questions:
\begin{itemize}
    \item \textbf{What are Blind and Low Vision people’s use cases for MLLM-enabled visual interpretation applications?}
    \item \textbf{How well are BLV people’s needs addressed in these identified use cases?}
    \item \textbf{How do Blind and Low Vision people converse with MLLM-enabled visual interpretation applications to address their needs?}
\end{itemize}

In this late-breaking work, we share preliminary answers to these research questions and discuss our plan to fully address these questions in future work. We conducted a two-week diary study followed by an interview with 20 BLV participants. During the diary study, participants used an MLLM-enabled visual interpretation application that we developed to emulate existing MLLM-enabled visual interpretation applications. However, it allowed participants to submit a short survey after each use. The survey, along with the user’s photo, the MLLM-generated description, and the conversation between the user and the system constituted a diary entry. We collected 553 diary entries and analyzed 60 entries in this preliminary analysis (10 randomly selected entries from six participants). 

We found that participants were very satisfied with most uses of the application, providing a mean satisfaction score of 4.15 out of 5 (SD=1.10), and they found the application very trustworthy, with a mean score of 3.75 out of 5 (SD=1.03). After receiving the initial interpretation, 75\% of diary entries included a follow-up conversation. Among those conversations, the mean number of messages was only 3.24 (SD=2.35, includes both user and AI messages), indicating that users tended to ask one or two questions. Participants' increased trust and satisfaction in the interactions led them to use our application in high-stakes scenarios, like identifying medication and receiving advice about dosages.

These results highlight that current MLLM-enabled visual interpretation can foster higher user satisfaction and trust, even in high-stakes scenarios. However, questions remain about how BLV users adapt their problem-solving strategies and navigate complex tasks with these systems in conversation. Building on our preliminary findings, we will complete an analysis of the full set of 553 diary entries to gain a deeper understanding of the impact of MLLMs in visual interpretation systems. We hope that these insights will help guide the creation of more robust, user-centered visual interpretation systems for Blind and Low Vision people, grounded in real people's experiences.

\section{Related Work}

We contribute to the body of knowledge exploring the use of AI-powered visual interpretation systems. Our work builds on prior research examining how BLV people interact with sighted collaborators, as well as studies investigating the use of AI-powered visual interpretation systems by BLV people.

\subsection{Human-Powered Visual Interpretation Systems}

Early research on visual interpretation for BLV people primarily focused on real-time navigational guidance rather than general visual interpretation tasks \cite{garaj2003system, bujacz2008remote}. For example, Garaj et al. created one of the first systems where a BLV user could share their point-of-view and localization with a remote sighted assistant to receive real-time navigational guidance \cite{garaj2003system}. Similarly, Bujacz et al. conducted a user study to evaluate the effects of in-person vs remote-sighted assistance in a comparable experimental setup \cite{bujacz2008remote}. Both studies found that remote-sighted assistance was useful to support BLV users’ navigation, although most participants still preferred in-person support.

Later, researchers introduced visual interpretation applications for general tasks. Applications such as VizWiz leveraged crowdsourced sighted volunteers to address BLV users’ visual information needs \cite{bigham-vizwiz}. After deploying VizWiz out in the wild, researchers were able to create a database of 72{,}205 photos and visual questions from BLV users \cite{gurari2018vizwiz}. While BLV users found this application helpful, the asynchronous nature of such systems render them unusable for visually complex tasks where real-time assistance is necessary. Moreover, users were constrained to a single query per photo to be answered by untrained human interpreters. Real-time assistance can allow both users and assistants to actively exchange information in real-time to achieve the BLV user’s goal.

Thus, visual interpretation applications such as Be My Eyes and AIRA were developed to connect BLV users with trained sighted assistants in real-time conversations \cite{nguyen2019large, lee2018conversations, xie-pair-volunteers}. In a large-scale assessment of 10{,}022 AIRA calls, researchers found that BLV users call remote sighted assistants to help with many daily activities, such as reading, social gatherings, shopping, and employment assistance \cite{nguyen2019large}. Other researchers explored how BLV people and remote sighted assistants interact to succeed in such visual interpretation tasks. They found that assistants probe BLV people to provide more personal knowledge to individually tailor their guidance as needed. Likewise, BLV people provide immediate feedback to remote sighted assistants, and also communicate in parallel through hand gestures for confirmation and navigation purposes \cite{lee2018conversations}. In an effort to improve remote-sighted assistance, Xie et al. proposed new modes of interaction to help remote sighted assistants guide users, including providing a real-time interactive 3D map of the BLV person’s location \cite{xie-helping-helpers}, and connecting multiple assistants at a time with one BLV person \cite{xie-pair-volunteers}. In this late-breaking work, we explored how BLV people interact with an MLLM-enabled application to answer visual questions in near real-time conversations.

\begin{table*}[ht]
    \caption{Participant demographics. The description of vision is self-reported by participants. }
    \centering
    \renewcommand{\arraystretch}{1.2}
    
    \begin{tabular}{|c|c|c|c|c|}
    \hline
        PID & Age & Gender & \parbox[c]{6cm}{\centering Description of Vision}   & Visual Interpretation Systems \\ \hline
        P1  & 41 & Male   & \centering Low Vision; Legally Blind; Peripheral and Central Vision Loss  & BeMyAI; SeeingAI; AIRA; Envision AI \\ \hline
        P2  & 32 & Female & \parbox[c]{6cm}{\centering Blind/No Light Perception}  & BeMyEyes/BeMyAI; SeeingAI; Envision AI \\ \hline
       P3 & 24 & Female & \centering Legally Blind; Color and Night Blindness; Peripheral Vision Loss  & BeMyAI; SeeingAI; AIRA; Envision AI \\ \hline
        P4 & 72 & Female & \parbox[c]{6cm}{\centering Blind/No Light Perception} & BeMyAI; SeeingAI; AIRA  \\ \hline
        P5 & 35 & Male   & \parbox[c]{6cm}{\centering Blind/Light Perception Only}  & BeMyAI; SeeingAI; AIRA; Speakaboo \\ \hline
        P6 & 44 & Male   & \parbox[c]{6cm}{\centering Blind/No Light Perception}  & BeMyAI; SeeingAI \\ \hline
    \end{tabular}
    
    \label{tab:participant-details}
\end{table*}
\subsection{AI-Powered Visual Interpretation Systems }
AI-powered systems, like SeeingAI, were the first tools that allowed BLV users to request visual descriptions without relying on sighted assistants (remote or otherwise). Users could choose to submit their photos for descriptions and access various features including reading text, identifying familiar people, scanning commercial products, and describing visual scenarios \cite{seeingAI}. This interaction paradigm afforded users to independently decide what content is described, when, and with what tools. Yet, in our prior work \cite{gonzalez2024usecases} we found that BLV users are frequently skeptical (2.43 on a 4-point trust scale SD=1.16) and unsatisfied (2.76 on a 5-point satisfaction scale SD=1.49) with AI-powered visual interpretations due to frequent errors in their descriptions. Similar work by \citet{kupferstein2020understanding} found low levels of satisfaction and trust too.

As AI has progressed, applications like Be My AI and SeeingAI have been updated to use MLLMs, such as GPT-4V \cite{be_my_ai_blog, applevis2025seeingai}. Compared to prior computer vision models, MLLMs have a wider range of knowledge and superior understanding of visual representations, leading to greater capabilities in generating detailed visual descriptions \cite{carolan2024review}. Because MLLMs have accelerated the development of conversational user interfaces, researchers have begun exploring the efficacy of MLLMs answers for user-generated visual questions \cite{longvqa_huh, chen2025fully}. For example, \citet{chen2025fully} collected an extensive dataset of user-submitted visual questions from online forums to evaluate multiple models visual answering performance. They found that, despite multiple models' abilities to create well-written long-form answers, they still performed poorly on topics with fewer existing datasets (e.g., religious-related topics and non-English languages). On the other hand, \citet{longvqa_huh} evaluated the performance of 6 MLLMs in providing long-form answers to a subset of pairs of visual questions and photos from the VizWiz dataset. They first generated the responses, and then subsequently asked 20 BLV people to observe and score the generated responses. Their participants scored the responses given a context prompt and used different metrics such as their plausibility and expected usefulness. They found that BLV participants were likely to think an answer is plausible as long as the answer is in long-form, regardless of its real correctness. These studies demonstrate the potential of MLLMs for answering visual questions from BLV people, but still do not provide any understanding of how BLV people are currently engaging with these applications to meet their daily needs.

As such, a group of researchers has explored how BLV people are adopting MLLM-enabled tools into their lives \cite{aigeneratedimg_bennett, contesting_alharbi, altextgenai_das, UseAi_das}. For example, Adnin et al. interviewed 19 Blind people to understand how and why Blind people access MLLM-enabled applications like ChatGPT, and Gemini. They found that Blind users leverage these applications to streamline their content creation workflows but face many accessibility barriers when navigating their interfaces. These works present recent accounts of BLV people’s usage of MLLM-enabled tools and provide critical considerations for designing future Human-AI interactions sensible to BLV people’s needs. We expand upon this body of work by examining how users interact, refine, and guide an MLLM-enabled visual interpretation application to meet their daily needs.

\section{Methods}

To understand how BLV people use MLLM-enabled visual interpretation applications, we conducted a diary study and interviews with BLV participants. We developed VisionPal, an MLLM-enabled visual interpretation application, that allowed participants to submit a short survey about their experience with every use. We recruited 20 participants who used VisionPal over the course of two weeks, and analyzed a random sample of 60 entries from 6 participants (10 entries per participant).

\subsection{Participants}

We recruited participants through a nonprofit organization serving Blind and Low Vision people. To be eligible for our study, participants had to be at least 18 years old, self-identify as Blind or Low Vision, and have an iOS device running iOS 15 (or later) to install and use our application. Upon completion of the study, participants received \$100 US dollars in compensation through an online platform. Participants had a choice to receive an electronic payment by gift cards, Venmo, or direct bank deposit. The study was approved by our university’s institutional review board. 

For this preliminary analysis, we randomly selected six participants out of 20 who participated in the study. Five of these participants were located in the United States, and the remaining participant was located in Turkey. Three were male and three were female, with ages ranging from 24 to 72 years (mean=41.3, SD=16.6). Five identified as Blind, and one participant identified as Low Vision. Demographics are listed in further detail in Table \ref{tab:participant-details}.

\subsection{Procedure}
\begin{figure*}
  \includegraphics[width=\linewidth]{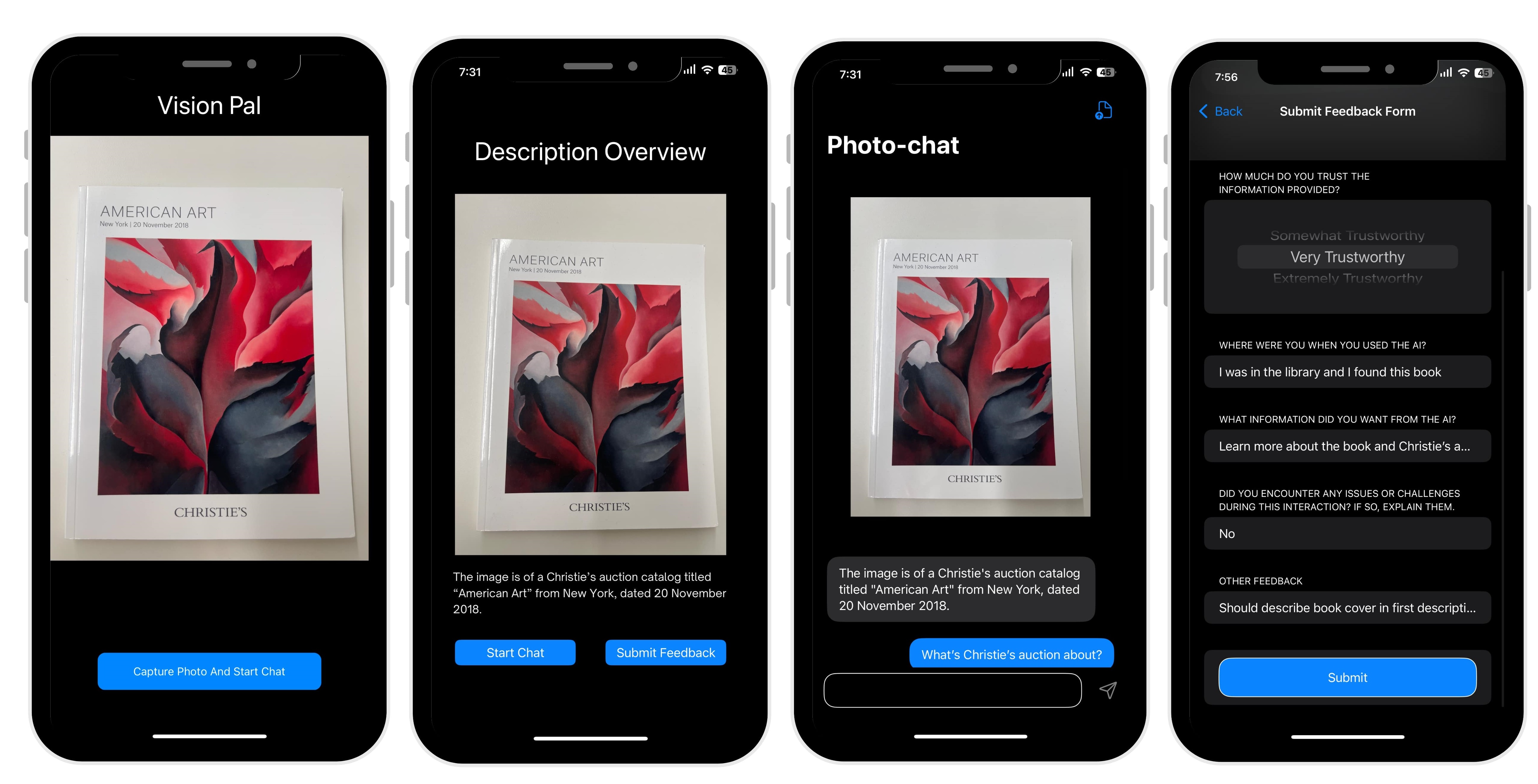}
  \caption{VisionPal, the visual interpretation application used to collect data. The screenshots illustrate the flow of submitting a diary entry: (1) photo capture, (2) receive initial photo description, (3) initiate conversation, and (4) submit feedback form.}
  \Description{VisionPal, the visual interpretation application used to collect data. The screenshots illustrate the flow of submitting a diary entry: (1) The first photo shows a phone with the first screen of the interface and the user is capturing a photo of a book called "American Art" and the interface has only 1 button that says "Capture Photo And Start Chat", (2) The second photo shows a second screen with the title "Description Overview", under the title the image the user took of the book, under that a short description reading "The image is of a Christie's auction catalog titled "American Art" from New York, dated 20 November 2018" and two buttons (one to start a chat and another to move to the short survey), (3) The third photo shows a typical chat interface where the user asked the question "What's Christie's auction about" and no visible response, and (4) The fourth and final image shows a short survey with five questions visible and a button to submit the survey. The first question visible is a picker item where the user is asked "how much do you trust the information provided?" and then the remaining questions are open-ended, "where were you when you used the AI?", "what information did you want from the ai?", "Did you encounter any issues or challenges during this interaction? If so explain", and "other feedback".}
  \label{fig:teaser}
\end{figure*}

The procedure of this study was similar to that of our previous diary study \cite{gonzalez2024usecases}.
The study consisted of an initial interview, a two-week diary period, and a follow-up interview. 

The initial interview was conducted through a video-conferencing platform and lasted approximately 60 minutes. During the initial session, we asked participants questions about their visual condition and their current use of AI-powered visual interpretation applications for meeting daily needs. Then, we instructed participants on using and submitting a diary entry with VisionPal. Finally, participants used VisionPal and submitted a diary entry on their own to verify they were able to independently use the application.

During the two-week diary period, participants were asked to use the application at least once a day. A diary entry included the photo taken, the initial returned visual description, the chat messages between the participant and the MLLM, and participants' responses to a short survey. Our short survey prompted participants to share information about their interaction with the MLLM, and we included six questions: 

\begin{itemize}
    \item How satisfied were you with the information provided? (Very dissatisfied to very satisfied)
    \item How much do you trust the information provided? (Not at all trustworthy to extremely trustworthy)
    \item Where were you when you used the AI?
    \item What information did you want from the AI?
    \item Did you encounter any issues or challenges during this interaction?
    \item Add any additional feedback.
\end{itemize}

At the end of the two-week diary period, we conducted a follow-up semi-structured interview via a video-conferencing platform, lasting 60 to 90 minutes. The purpose of this interview was to understand participants' overall experiences interacting with the MLLM and to gain deeper insights into diary entries that showed unexpected use cases.

We divided the follow-up interviews into three parts: questions to gather overall impressions about participants' experiences with the application; questions to probe for more context regarding selected diary entries; and questions to rate their overall experience using VisionPal.

\subsection{ VisionPal: Our Data Collection Application}

The goal of our study was to understand how Blind and Low Vision people use MLLM-enabled visual interpretation applications in general. To capture usage of visual interpretation applications that mirror current BLV users’ experiences, we developed our own application, as mainstream services (e.g., SeeingAI, Be My AI) were not open source and did not offer mechanisms for researchers to collect usage data. Moreover, prior work has reported significant challenges with participant attrition and receiving lower quality data \cite{kupferstein2020understanding} when participants manually self-report their interactions with visual interpretation applications (e.g., emailing summaries of their interactions with the system to researchers).

Then, following our methodology in prior work to collect high-quality data in diary studies \cite{gonzalez2024usecases}, we developed a custom application that offers a similar user experience to mainstream commercial applications (e.g., SeeingAI, Be My AI), ensuring an accessible process for submitting diary entries. This approach allowed us to capture participants’ feedback immediately after they interacted with the system. We chose the GPT-4o model from OpenAI to generate our visual interpretations because GPT-4V, its predecessor, has shown strong performance for answering visual questions \cite{longvqa_huh, openai_gpt4v_system_card}. The GPT-4V family of models offers a broad range of capabilities such as optical character recognition, abstract visual reasoning, emotional awareness, scene understanding, and other capabilities, all in a single model \cite{yang2023dawn}. Furthermore, Be My AI and SeeingAI are known to be powered by GPT-4 family models, or equally advanced models to generate their visual interpretations \cite{openai_gpt4v_system_card, seeingaigpt4}. Thus, GPT-4o was one of the best choices for creating a conversational visual interpretation application that is representative of mainstream visual interpretation applications.

Our application, VisionPal, closely followed the design and interaction flow of Be My AI and SeeingAI. In these applications, BLV users can take a photo, receive a visual description of the image, open a conversation to explore the image further, and then provide feedback or close the conversation before starting a new interaction. In VisionPal, participants can open the application and immediately take a photo (Figure \ref{fig:teaser}) to receive an initial description. Then, participants have the option to either open a chat for follow-up questions or proceed to a short survey if they are done with the interaction. If they opened the chat, they accessed an interface where they could exchange messages with the MLLM. Each time participants sent a new message, the system executed an API call to the GPT-4o model with the entire conversation and photo as context to generate responses. When participants were finished chatting, they filled out a six question short survey and submitted the diary entry.

\subsection{Preliminary Analysis}

We collected 553 diary entries from our 20 participants. For this preliminary analysis, we randomly selected six participants to sample their diary entries. For each participant, we used a uniform random approach to select 10 diary entries (e.g., if a participant had 34 entries, we randomly sampled 10 distinct entries with all entries having equal probability to be selected). This resulted in a total of 60 diary entries from our six participants. 

To understand how participants interact with MLLM-enabled visual interpretation applications, we focused our analysis on the conversations of the diary entries. These conversations revealed how participants interact with the MLLM, and what kinds of questions they asked. Moreover, conversations showed why, when, and how MLLM visual interpretation applications fail to meet user expectations and needs. We supplied this analysis with participants' commentary from the follow-up interviews.

To analyze the conversations, we applied two groups of closed codes. One group of closed codes was adopted from the taxonomy proposed by Chen et al. to identify answer-seeking behavior from our users \cite{chen2025fully}. The second group of closed codes identified all questions participants asked and whether the questions were answered correctly, partially, incorrectly, or left unanswered (This was determined by the visual context, and whether the system abstained as needed). Two researchers coded 20 conversations individually, and then discussed disparities to ensure consistency in coding practice. Then, the remaining 40 conversations were coded by one researcher.

For the follow-up interviews, we used an inductive open coding process. Two researchers coded two interviews individually and discussed disparities to develop a codebook. Finally, the remaining four interviews were coded by one researcher.

Upon completing these coding processes, one researcher conducted two rounds of affinity diagramming to develop preliminary themes which, subsequently, were validated by four researchers.

\section{Findings}

We present findings on participants’ satisfaction and trust, their conversational interactions, and the types of questions they asked, illustrating how they engaged with our MLLM-enabled visual interpretation application.

\subsection{Satisfaction and Trust}

Of our 60 diary entries, participants submitted 48 entries (80\%) with the survey. The mean score for trust across all diary entries was 3.75 out of 5 (SD=1.03), where 5 meant participants thought the information provided by the AI was extremely trustworthy. The mean score for satisfaction was 4.15 out of 5 (SD=1.10), where 5 meant participants thought the information provided by the system was very satisfying. This indicates participants were very likely to be satisfied with the information received, but remained slightly skeptical of the AI. P3 explained that they “never fully trusted the app” because “One, it’s just a personal policy and…it has just been my experience that still, human-generated descriptions are 99\% of the time more accurate.” While participants were encouraged to submit the survey for all entries, participants occasionally closed the application immediately after receiving the initial description or experienced API connection timeouts, leading to 12 entries (20\%) with incomplete survey submissions.

\subsection{Conversations }

In total, there were 45 diary entries (75\%) where participants started a conversation. While most of our participants had a maximum conversation length of two messages (one by the user and one by the AI), some participants extensively engaged with the AI (up to 14 messages, seven by the user, and seven by the AI). The average length of conversations by participant ranged from 2 to 6.89, with a total average of 3.24 (SD=2.35) across all six participants. For the remaining 15 entries (25\%), participants immediately submitted feedback in 3 entries (5\%) or they immediately closed the application after receiving a description or an error from the system in 12 entries (20\%). 

\begin{table}[h!]
  \caption{Summary of conversation metrics by participant. Some participants engaged with the AI extensively, while others sent and received at most one follow-up message (one by the user, and one by the AI).}
  \centering
  \renewcommand{\arraystretch}{1.2}
  \begin{tabular}{|p{0.6cm}|p{2cm}|p{2cm}|p{2cm}|}
    \hline
    \parbox[c][6.5ex][c]{0.6cm}{\centering PID} & 
    \parbox[c][6.5ex][c]{2cm}{\centering Conversations} & 
    \parbox[c][6.5ex][c]{2cm}{\centering Max.\ Length of Conversations} & 
    \parbox[c][6.5ex][c]{2cm}{\centering Avg.\ Length of Conversations} \\ \hline
    \parbox[c]{0.6cm}{\centering P1}  & \parbox[c]{2cm}{\centering 9}  & \parbox[c]{2cm}{\centering 14} & \parbox[c]{2cm}{\centering 6.89} \\ \hline
    \parbox[c]{0.6cm}{\centering P2}  & \parbox[c]{2cm}{\centering 6}  & \parbox[c]{2cm}{\centering 2}  & \parbox[c]{2cm}{\centering 2} \\ \hline
    \parbox[c]{0.6cm}{\centering P3}  & \parbox[c]{2cm}{\centering 7}  & \parbox[c]{2cm}{\centering 2}  & \parbox[c]{2cm}{\centering 2} \\ \hline
    \parbox[c]{0.6cm}{\centering P4}  & \parbox[c]{2cm}{\centering 8}  & \parbox[c]{2cm}{\centering 2}  & \parbox[c]{2cm}{\centering 2} \\ \hline
    \parbox[c]{0.6cm}{\centering P5}  & \parbox[c]{2cm}{\centering 5}  & \parbox[c]{2cm}{\centering 4}  & \parbox[c]{2cm}{\centering 3} \\ \hline
    \parbox[c]{0.6cm}{\centering P6}  & \parbox[c]{2cm}{\centering 10} & \parbox[c]{2cm}{\centering 4}  & \parbox[c]{2cm}{\centering 2} \\ \hline
  \end{tabular}
  \label{tab:conversation-details}
\end{table}

\begin{table*}[ht]
  \centering
  \caption{Summary of types of questions by participant. The majority of questions were related to identifying or verifying something visually, or asking evidence-based questions.}
  \label{tab:types_of_questions}
  \renewcommand{\arraystretch}{1.2}
  \setlength{\tabcolsep}{6pt}
  \begin{tabular}{|c|p{1.8cm}|p{2cm}|p{1.8cm}|p{1.8cm}|p{1.6cm}|p{1.4cm}|p{1.6cm}|}
    \hline
PID 
  & Evidence Questions
  & {\centering Identification Questions}
  & {\centering Verification Questions}
  & {\centering Instruction Requests}
  & {\centering Opinion Questions}
  & {\centering Advice Questions}
  & {\centering Reason Questions} \\
\hline

    P1
      & \parbox[c]{1.8cm}{\centering 14}
      & \parbox[c]{2cm}{\centering 7}
      & \parbox[c]{1.8cm}{\centering 7}
      & \parbox[c]{1.8cm}{\centering 2}
      & \parbox[c]{1.6cm}{\centering 2}
      & \parbox[c]{1.4cm}{\centering 0}
      & \parbox[c]{1.6cm}{\centering 2} \\
    \hline

    P2
      & \parbox[c]{1.8cm}{\centering 4}
      & \parbox[c]{2cm}{\centering 1}
      & \parbox[c]{1.8cm}{\centering 0}
      & \parbox[c]{1.8cm}{\centering 0}
      & \parbox[c]{1.6cm}{\centering 0}
      & \parbox[c]{1.4cm}{\centering 0}
      & \parbox[c]{1.6cm}{\centering 0} \\
    \hline

    P3
      & \parbox[c]{1.8cm}{\centering 5}
      & \parbox[c]{2cm}{\centering 2}
      & \parbox[c]{1.8cm}{\centering 1}
      & \parbox[c]{1.8cm}{\centering 0}
      & \parbox[c]{1.6cm}{\centering 0}
      & \parbox[c]{1.4cm}{\centering 0}
      & \parbox[c]{1.6cm}{\centering 0} \\
    \hline

    P4
      & \parbox[c]{1.8cm}{\centering 5}
      & \parbox[c]{2cm}{\centering 1}
      & \parbox[c]{1.8cm}{\centering 2}
      & \parbox[c]{1.8cm}{\centering 0}
      & \parbox[c]{1.6cm}{\centering 0}
      & \parbox[c]{1.4cm}{\centering 1}
      & \parbox[c]{1.6cm}{\centering 0} \\
    \hline

    P5
      & \parbox[c]{1.8cm}{\centering 6}
      & \parbox[c]{2cm}{\centering 3}
      & \parbox[c]{1.8cm}{\centering 3}
      & \parbox[c]{1.8cm}{\centering 0}
      & \parbox[c]{1.6cm}{\centering 0}
      & \parbox[c]{1.4cm}{\centering 0}
      & \parbox[c]{1.6cm}{\centering 0} \\
    \hline

    P6
      & \parbox[c]{1.8cm}{\centering 7}
      & \parbox[c]{2cm}{\centering 3}
      & \parbox[c]{1.8cm}{\centering 1}
      & \parbox[c]{1.8cm}{\centering 0}
      & \parbox[c]{1.6cm}{\centering 2}
      & \parbox[c]{1.4cm}{\centering 0}
      & \parbox[c]{1.6cm}{\centering 0} \\
    \hline

    Total
      & \parbox[c]{1.8cm}{\centering 41}
      & \parbox[c]{2cm}{\centering 17}
      & \parbox[c]{1.8cm}{\centering 14}
      & \parbox[c]{1.8cm}{\centering 2}
      & \parbox[c]{1.6cm}{\centering 3}
      & \parbox[c]{1.4cm}{\centering 1}
      & \parbox[c]{1.6cm}{\centering 2} \\
    \hline

  \end{tabular}
\end{table*}

Depending on the complexity of their use case, most participants received the information they desired within one or two messages. Otherwise, they engaged the AI in a longer conversation. For instance, when P1 was setting up his new braille embosser, the MLLM hallucinated braille labels while providing instructions to P1. He “tried to get [the AI] to tell me where the cartridges were [but] it failed spectacularly. It tried to tell me there were braille labels on the ink cartridges. That was crazy.” P1 often felt the need to challenge and guide the AI to correct its answers and future behavior, concluding that “AI still needs a lot of guidance from humans.” This highlights how participants' use cases have evolved to demand accurate domain knowledge (visual assistance) especially relevant to BLV people and to present it in a way that is consistent with its purpose of assisting BLV people address their needs independently.

\subsection{ Questions and Messages}
\begin{table}[ht]
  \caption{Summary of questions by participant. Most participants had a relatively consistent experience except P1 and P5, who did not receive some answers or received many incorrect and partial answers.}
  \label{tab:questions_answered}
  \centering
  \renewcommand{\arraystretch}{1.2}
  \begin{tabular}{|p{0.5cm}|p{1.2cm}|p{1.15cm}|p{1.15cm}|p{1.15cm}|p{1.15cm}|}
    \hline
    PID & Questions & Correct Answers & Incorrect Answers & Partial Answers & Un-answered \\ \hline
    \parbox[c]{0.5cm}{\centering P1}  & \parbox[c]{1.2cm}{\centering 33}  & \parbox[c]{1.15cm}{\centering 23}  & \parbox[c]{1.15cm}{\centering 5}   & \parbox[c]{1.15cm}{\centering 3}  & \parbox[c]{1.15cm}{\centering 2}  \\ \hline
    \parbox[c]{0.5cm}{\centering P2}  & \parbox[c]{1.2cm}{\centering 6}   & \parbox[c]{1.15cm}{\centering 6}   & \parbox[c]{1.15cm}{\centering 0}   & \parbox[c]{1.15cm}{\centering 0}  & \parbox[c]{1.15cm}{\centering 0}  \\ \hline
    \parbox[c]{0.5cm}{\centering P3}  & \parbox[c]{1.2cm}{\centering 8}   & \parbox[c]{1.15cm}{\centering 5}   & \parbox[c]{1.15cm}{\centering 2}   & \parbox[c]{1.15cm}{\centering 1}  & \parbox[c]{1.15cm}{\centering 0}  \\ \hline
    \parbox[c]{0.5cm}{\centering P4}  & \parbox[c]{1.2cm}{\centering 9}   & \parbox[c]{1.15cm}{\centering 7}   & \parbox[c]{1.15cm}{\centering 1}   & \parbox[c]{1.15cm}{\centering 1}  & \parbox[c]{1.15cm}{\centering 0}  \\ \hline
    \parbox[c]{0.5cm}{\centering P5}  & \parbox[c]{1.2cm}{\centering 11}  & \parbox[c]{1.15cm}{\centering 7}   & \parbox[c]{1.15cm}{\centering 4}   & \parbox[c]{1.15cm}{\centering 0}  & \parbox[c]{1.15cm}{\centering 0}  \\ \hline
    \parbox[c]{0.5cm}{\centering P6}  & \parbox[c]{1.2cm}{\centering 13}  & \parbox[c]{1.15cm}{\centering 10}  & \parbox[c]{1.15cm}{\centering 3}   & \parbox[c]{1.15cm}{\centering 0}  & \parbox[c]{1.15cm}{\centering 0}  \\ \hline
    \parbox[c]{0.5cm}{\centering Tot.}  & \parbox[c]{1.2cm}{\centering 80}  & \parbox[c]{1.15cm}{\centering 58}  & \parbox[c]{1.15cm}{\centering 15}  & \parbox[c]{1.15cm}{\centering 5}  & \parbox[c]{1.15cm}{\centering 2}  \\ \hline
  \end{tabular}
\end{table}

Our participants asked 80 questions in total (See Table \ref{tab:questions_answered}). Fifty-eight questions (72.5\%) were answered correctly, 15 questions (18.75\%) were answered incorrectly, 5 questions (6.25\%) were answered partially, and 2 questions (2.5\%) were left unanswered. Questions were often asked when the initial description failed to provide enough detail, such as: “Tell me what brand and types of chips are in the visible bags” (P1). In other situations, participants would ask speculative questions to the AI, like when P6 asked, “From which regeon [sic] of the world [this rug] could be from?” Similarly, P4 captured a photo of a textbook and later asked the AI, “What is this book about what is is [sic] there more information?”

In Table \ref{tab:types_of_questions}, we present the types of questions participants asked. Forty-one questions were evidence-based questions (51.25\%), such as an image of a nutritional label submitted by P4, followed by the question: “Does this bread have fiber and how much?” Other common types of questions were to identify (17 times, 21.25\%) or verify (14 times, 17.5\%) something visually, like: “Do you know which [piano] model specifically?” For more complex queries, like instruction, opinion, advice and reason questions (why questions), participants only asked 8 questions (10\%). For example, P4 sought assistance from the system to learn what is the recommended dosage for a medication she needed: “What amount of [Tylenol] medication should I take?” While low in count, these questions demonstrate users’ evolving information-seeking needs and their future expectations for AI support.

\section{Discussion and Future Work}
Since MLLMs represent a significant advancement in AI, we sought to understand BLV people’s use of MLLM-enabled visual interpretation applications. We conducted a similar study to our previous work \cite{gonzalez2024usecases}, which was conducted before MLLMs were incorporated into these applications. Based on this prior work, we posed three research questions: in what use cases do BLV people use MLLM-enabled visual interpretation applications, how well are BLV people's needs met for those use cases, and how do they converse with the MLLMs in these applications? In this late-breaking work, we presented a preliminary analysis of a subset of our data, 60 diary entries from six of the 20 BLV participants in our study. 

We found that participants were satisfied with the vast majority of their interactions with the MLLM-enabled application, and they leaned towards trusting the information they received. This contrasted with our previous work \cite{gonzalez2024usecases}, which reported many instances of dissatisfaction and distrust among participants. The mean score for satisfaction across diary entries was 4.15 out of 5 (SD=1.10) in the current study, and 2.76 out of 5 (SD=1.49) in our previous study \cite{gonzalez2024usecases}; meanwhile, the mean score for trust in the current study was 3.75 out of 5 (SD=1.03) and 2.43 out of 4 (SD=1.16) in our previous study \cite{gonzalez2024usecases}. In the current study, participants used the application in high-stakes and complex use cases, such as receiving medical dosage information or setting up a braille embossing machine. This suggests that AI-powered visual interpretation applications have reached a threshold of usability and trustworthiness to support BLV people with many daily life tasks. 

However, these initial findings pose a new question: despite frequent satisfaction and trust with information received, were these positive impressions reflective of reality? In fifteen diary entries (18.75\%), the application provided incorrect information, in five entries (6.25\%) it provided partially correct information, and in two entries (2.5\%) it did not provide any information to address the user’s question. This indicates that only measuring and optimizing for user satisfaction and trust may mislead or harm users in unintended ways. Both \citet{UseAi_das} and \citet{longvqa_huh} found something similar in their studies, where BLV participants reported high levels of trust in detailed descriptions and answers regardless of their true correctness. As MLLMs continue to improve their abilities and become more integrated into everyday life, it is important to capture the context in which these errors happen to proactively improve visual interpretation systems and set users’ expectations appropriately. For example, P1’s failed attempt to set up a braille embosser shows that MLLMs still face significant challenges communicating appropriately with BLV users (e.g., hallucinated nonexistent braille labels) and providing accurate information for input that is not well-represented in training datasets \cite{chen2025fully}.

Given these findings, it is increasingly important to gather up-to-date data on BLV people’s use of MLLM-enabled visual interpretation applications, especially as AI-powered assistants become more widely adopted and transition to wearable formats, such as Meta’s Ray-Ban AI glasses \cite{rayban}, Project Aria \cite{engel2023projectarianewtool} and Google’s Project Astra \cite{projectastra} smart glasses. By studying these authentic conversations, we can better understand how new conversational modalities shape the questions BLV users ask, the clarifications they require, and the level of trust they develop in AI-generated responses and AI assistants over time. 

Our analysis of 60 diary entries provides preliminary answers to address our three research questions. However, our larger dataset of 553 entries will offer deeper insights into how BLV users’ interactions with AI-powered visual interpretation applications have evolved their information-seeking behaviors and impacted their daily lives. Furthermore, our analysis was limited because we only analyzed randomly sampled diary entries, rather than analyzing participants' complete diaries. In future work, we plan to conduct a more comprehensive analysis of the remaining diary entries in our dataset to fully identify current use cases for MLLM-enabled visual interpretation applications, understand users’ patterns of use over time, and evaluate how well these MLLMs address users' needs. To do this, we will expand our qualitative analysis by applying qualitative codes on diary entries to identify our participants’ goals for each conversation, location of use, and the accuracy of the AI-generated outputs. We will also complement these new applied codes with statistical tests to evaluate whether the MLLM performed significantly better for specific use cases or more accurately met particular user goals. These insights will help guide the creation of more robust, user-centered visual interpretation systems for Blind and Low Vision people, grounded in real people's experiences.

\bibliographystyle{ACM-Reference-Format}


\end{document}